\documentclass[conference]{IEEEtran}
\IEEEoverridecommandlockouts
\usepackage{cite}
\usepackage{amsmath,amssymb,amsfonts}
\usepackage{algorithmic}
\usepackage{graphicx}
\usepackage{textcomp}
\usepackage{xcolor}
\usepackage{hyperref}
\def\BibTeX{{\rm B\kern-.05em{\sc i\kern-.025em b}\kern-.08em
    T\kern-.1667em\lower.7ex\hbox{E}\kern-.125emX}}
\begin{document}

\title{Turning Software Engineers into AI Engineers\\
\thanks{This work is financed by the Taskforce for Applied Research (SIA), part of the Netherlands Organisation for Scientific Research (NWO)}
}

\author{\IEEEauthorblockN{Petra Heck}
\IEEEauthorblockA{\textit{Fontys University of Applied Sciences} \\
Eindhoven, Netherlands \\
p.heck@fontys.nl}
\and
\IEEEauthorblockN{Gerard Schouten}
\IEEEauthorblockA{\textit{Fontys University of Applied Sciences} \\
Eindhoven, Netherlands \\
g.schouten@fontys.nl}
}

\maketitle

\begin{abstract}
In industry as well as education as well as academics we see a growing need for knowledge on how to apply machine learning in software applications. With the educational programme ICT \& AI at Fontys UAS we had to find an answer to the question: \textit{How should we educate software engineers to become AI engineers?} This paper describes our educational programme, the open source tools we use, and the literature it is based on. After three years of experience, we present our lessons learned for both educational institutions and software engineers in practice.   
\end{abstract}


\section{Introduction}
Fontys University of Applied Sciences (Fontys UAS) is a higher education institution with profession-oriented study programmes. The final semester of each ICT (Information and Communication Technology) programme at Fontys UAS consists of a practical assignment at an external company. More and more of these practical assignments have a machine learning component in them. We see that machine learning has become available outside an academic setting as a new technology for engineers to improve processes, services and products. This led us to the question: how should we educate our students to become machine learning engineers?

At the same time society reflects and acts on the world-wide Artificial Intelligence (AI) trend, trying to formulate answers on how to deal with AI. Newspapers write about AI, governments are starting local and national AI programmes, conferences are organised, etc. From an educational perspective this means we have to make sure that our ICT students know what AI is and how they should use it to build ICT applications. The world-wide hype on AI made us slightly rephrase our question to: how should we educate our students to become AI engineers? Note that with the current state of practice AI engineering is synonym with Machine Learning (ML) engineering but the term AI corresponds better to the societal view.

For the engineering part of the question at Fontys UAS we have a long time experience in educating software engineers. Our yearly quality control cycle with industry partners ensures that the educational programme is kept up-to-date and addresses the needs of the (regional) software engineering industry. In general, they are more than happy with both the technical and professional skills of our ICT bachelor students. However, the AI part of the question turned out to be new, even for the software engineering community\cite{Menzies,Bosch}. There is a plethora of machine learning material to be found online but it is difficult to know where to start when you need to apply AI technology in practical projects. This discovery in the software engineering industry made us change our question to a broader one: how should we educate software engineers to become AI engineers? Answers we find for our students are also valid for software engineers in industry that need to switch to AI applications. 

In 2017, anticipating on the AI trend, we added a specialisation programme in AI to our regular Software Engineering programme. The contents of the AI programme are determined by our continuous study of the state-of-the-practice for machine learning as well as our hands-on educational philosophy. We try to give a broad overview of the machine learning field, letting the students go in-depth on the topics of their choice. This AI programme is our answer to the question: how should we educate software engineers to become AI engineers. 

In this paper, we describe our educational programme in terms of semesters, learning outcomes, high-level content and student activities. We have conducted informal interviews and a questionnaire with a small group of students to collect initial feedback on the first version of our programme. We summarise the changes planned for the second version of our programme. We end the paper by discussing our programme and the feedback we received. We highlight the most important findings in the form of \textbf{lessons learned}, which are also useful for solving practical questions in AI projects in industry. 

Our finding is that the group of students that followed the complete AI programme is apt for their graduation assignment in industry. As we see practitioners moving their focus to the engineering part (instead of the modelling part) of machine learning, additional material around this (deployment, testing, tooling) is the major update we did for the second version of the programme. We will keep following the state-of-the-practice around AI engineering and keep updating our educational programme based on that. At the same time we will keep publishing these updates as best practices for practitioners.  

\section{Software Eng. \& AI programme at Fontys UAS}
\begin{figure}
\includegraphics[width=\columnwidth]{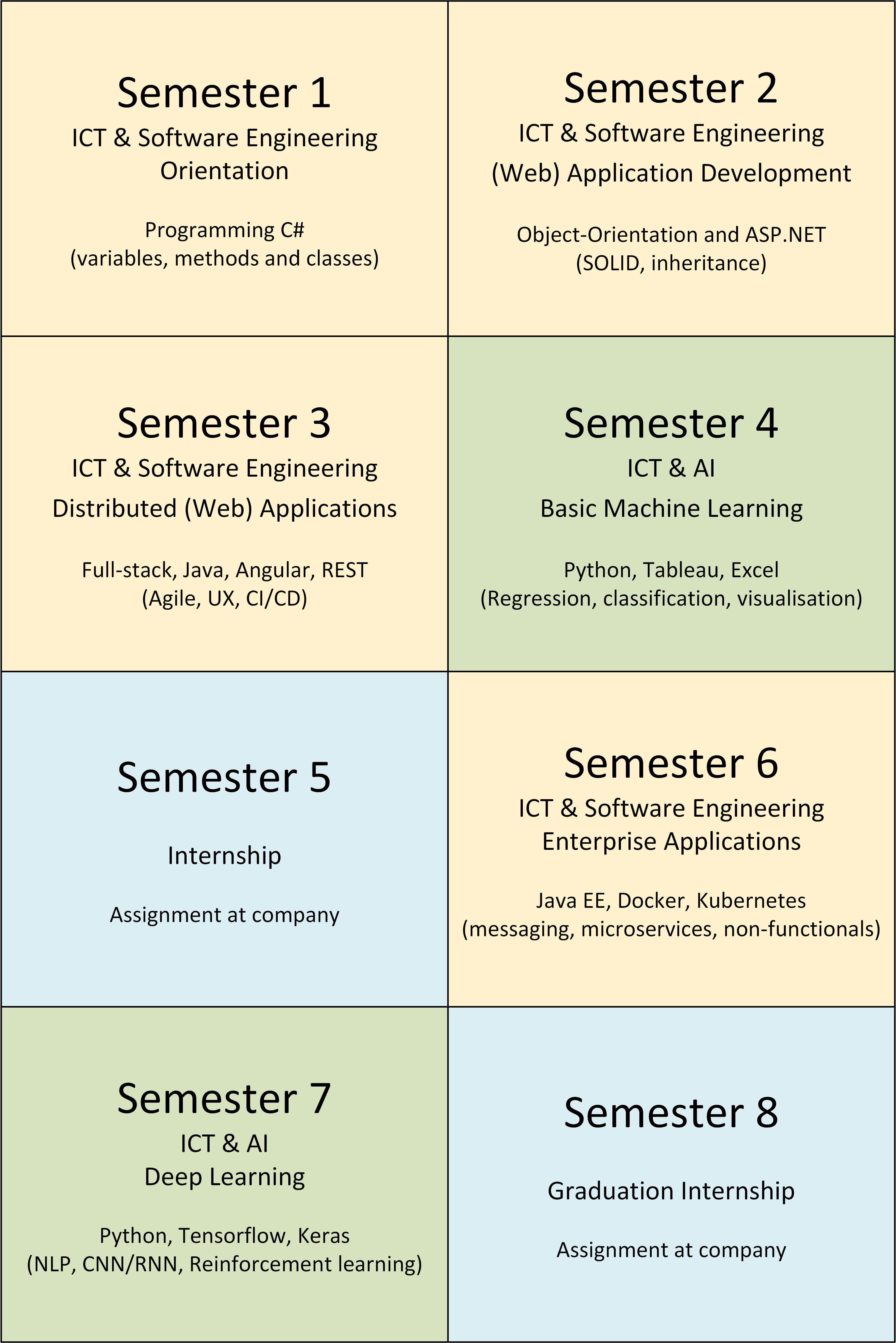}
\caption{Software Engineering \& AI programme at Fontys UAS}
\label{fig_curAI}
\end{figure}

ICT \& Software Engineering is a four-year programme that educates students for the profession of software engineer and exists for more than ten years. Since September 2017 students have the option to specialise in ICT \& AI. Such a combined programme (see Figure \ref{fig_curAI}) consists of: 
\begin{enumerate}
    \item four semesters of applied Software Engineering
    \item two semesters of applied AI 
    \item two internships engineering AI applications at and for an external organisation
\end{enumerate}
In June 2020, the first group of students that followed this complete programme graduated.

The content of each semester is provided to the students in an open way: we define high-level learning outcomes, provide example implementations of those learning outcomes, but let the students discover their own implementations. The students have to convince the teachers that they master the learning outcomes by applying newly acquired skills to their own assignments. Each semester is based around a case that comes from an external client (company or other organisation) for which we as teachers do not know beforehand what the outcome should be. We mentor the students as coach and expert to guide them to a working solution.

By combining the machine learning oriented content of ICT \& AI with the software engineering content of ICT \& Software Engineering we have a unique way of educating students for the profession of machine learning engineer. We realise that a similar modular approach also could work for software engineering practitioners that need to switch to engineering machine learning applications. That is why we want to share the content of our AI curriculum and the lessons learned for both educational institutions and practitioners.  

\section{ICT \& AI (Semester 4 and 7)}
In this section, we present the programme of the two semesters ICT \& AI. Both semesters are credited with 30 ECTS credits. 

\subsection{Learning Outcomes}
The set of learning outcomes for Semester 7 ICT \& AI is:
\begin{enumerate}
    \item The student is able to prepare and store a dataset in such a way that it can be used for applying data science techniques.
    \item The student can reliably apply machine (including deep) learning algorithms to a given dataset, creating an optimal model for the goal at hand.
    \item The student proposes a data-driven innovation with a viable business case for machine learning and is able to interact with the client about it in a comprehensible way.
    \item The student uses domain knowledge to develop relevant data-driven innovations following a structured approach.
    \item The student is an effective co-worker.
    \item The student guides his/her own study progress by asking for, interpreting and applying feedback from teachers, tutors, coaches and fellow students.
\end{enumerate}
The learning outcomes of Semester 4 ICT \& AI are similar to this but cover only the basic machine learning techniques (not deep learning). Of course the complexity level at which the learning outcomes have to be mastered increases between Semester 4 and Semester 7. For instance, whereas in Semester 4 students have to deal with a single project and only one stakeholder they have to run their own mini-company in Semester 7 and deal with several projects (typically two or three) in a multi-stakeholder setting. \\ 

\subsection{Tools}
In principle the students are free to use whatever tools they want to achieve the learning outcomes during the semester. For our teacher material, however, we chose one tool per activity:
\begin{itemize}
    \item Python with libraries pandas, matplotlib, scikit-learn for programming data wrangling, data visualisation and machine learning respectively.
    \item Jupyter notebooks as a programming environment
    \item Tensorflow 2.0 for programming deep learning
    \item Excel for interactive data processing (without programming)
    \item Tableau for interactive data visualisation (without programming)
\end{itemize}

\subsection{Assignments and Grading}
\begin{figure}
\includegraphics[width=\columnwidth]{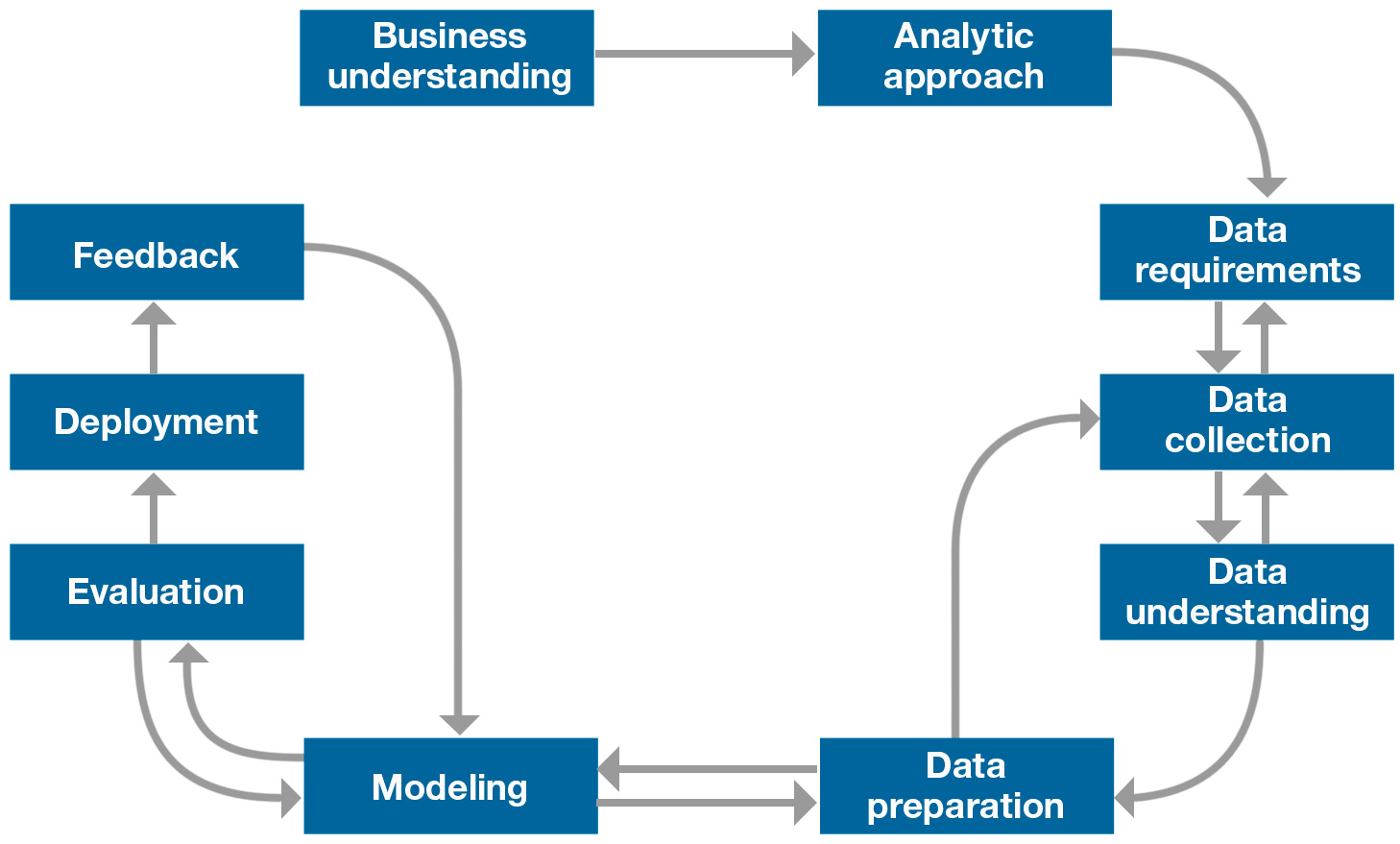}
\caption{IBM Data Science Methodology\cite{IBM}}
\label{fig_IBM}
\end{figure}

Both semesters have a similar structure of assignments that students have to hand in\footnote{We use the \href{instructure.com}{Canvas} online learning environment: one course per semester for all materials and assignments, feedback and grading}:
\begin{enumerate}
    \item\textit{Workshop Exercises} These are optional exercises to practice skills in isolation like cleaning a dirty dataset or applying a CNN to a given dataset or visualising a data set. The teacher gives a short introduction to each workshop topic. Exercises are not graded, but sometimes require teacher feedback.
    \item\textit{Challenges} These are mandatory individual assignments to integrate several skills. An example is to find your own dataset, clean it, choose a prediction target, predict using at least three different algorithms, and interpret the results using standard evaluation metrics like accuracy, recall and precision. The teacher grades the challenges (formative).
    \item\textit{Open Programme Report} Students are allowed to spend 10\% of their time on an AI-related topic of their own choice. This is because we know we can not cover all topics in just two semesters. It provides an elegant way of students teaching each other as they need to present findings. The teacher grades the open programme report (formative). 
    \item\textit{Group project} This is a project with an external stakeholder as a client where a group of students has to create a predictive model, starting from business requirements, going through data preparation, modelling and reporting. We use the IBM data science methodology \cite{IBM} to structure the project, see Figure \ref{fig_IBM}. The teacher mentors the group and grades this group assignment with input from the client (formative).
    \item\textit{Personal Development Report (PDR)} In the PDR students reflect on their progress towards the learning outcomes. They use teacher feedback to support this reflection. Based on the PDR (that links to all other assignments) the teachers determine the final grade for the semester for that individual student (summative).    
\end{enumerate}

This setup gives structure (workshops, group project) to the students while at the same time allowing them to choose their own way of proving the learning outcomes. During the semester they get formative grades and feedback; they just have to ensure that everything is ``repaired" by the end of the semester when they have their final summative assessment. The assignments are open such that students can deviate from the standard examples that we give in the exercises. They can e.g. use R instead of Python if they want, use different algorithms, choose their own dataset, integrate other data analysis tools and techniques. We challenge the students to go beyond the minimum requirements; especially in the group project where we ask them to come up with a solution that matches, or even better, exceeds the needs and expectations of the client.\\

\subsection{Content of Workshops}
The workshops in both semesters with accompanying exercises give students pointers to study material and skills they could use to fulfil the challenges and group project. Each workshop consists of an individual preparation, an introductory presentation by the teacher who also discusses the preparations, a group exercise that builds upon the individual preparation, and a group discussion of the results. We have workshops on the following topics (Semester 4 or 7 indicated by S4 or S7):
\begin{itemize}
    \item S4 Introduction Python and pandas
    \item S4 Data quality: a) data cleaning, b) handling missing data, c) reporting
    \item S4 Feature engineering
    \item S4 Basic statistics (mean, median, standard deviation, etc.)
    \item S4 Basic machine learning algorithms: a) regression, b) tree-based methods, c) SVMs, d) neural networks
    \item S4 Modelling techniques: a) cross-validation, b) model performance evaluation metrics, c) dimensionality reduction
    \item S4 Exploratory data visualisation with matplotlib
    \item S4 Data visualisation: a) basic charts; b) advanced analytics; c) visual design principles; d) Info-graphics; e) interactive visualisations 
    \item S4 Writing a business proposal
    \item S4 Ethical and legal aspects of AI projects
    \item S4 Organisational context: explaining the importance of data-driven maturity of an organisation when applying machine learning
    \item S4 Societal context: a) smart cities; b) sustainable development goals
    \item S4 Setting up a cloud environment: a) data sources; b) web scraping; c) cloud tools/APIs
    \item S7 Advanced machine learning algorithms: a) CNNs, b) RNNs, c) reinforcement learning
    \item S7 Natural Language Processing (NLP)
\end{itemize}

As one can see Semester 7 has less workshops. The students are more self-directed to collect the knowledge online they need to fulfil the client project and they build upon the workshops they received in Semester 4. In Semester 7 we reserve each Wednesday afternoon for inviting guest lecturers. This gives us the opportunity to highlight the state-of-the-practice not included in the base material of the semester yet and to enrich the theory with stories and examples from AI engineers in industry.\\

\section{Student Feedback on ICT \& AI Programme}
In September 2019, the first group of 25 students followed Semester 7. Out of those 25 students, nine students started their graduation project (Semester 8) ICT \& AI in February 2020. 

We interviewed the students twice: once during their minor about their group project (seven groups) and once during their individual graduation assignment (nine students). The type of interviews in the first round was semi-structured (informal setting with prepared questions by both authors, flexibility to probe the participants for additional details) and the second round was structured (online questionnaire, strict protocol).

\begin{figure}
\includegraphics[width=\columnwidth]{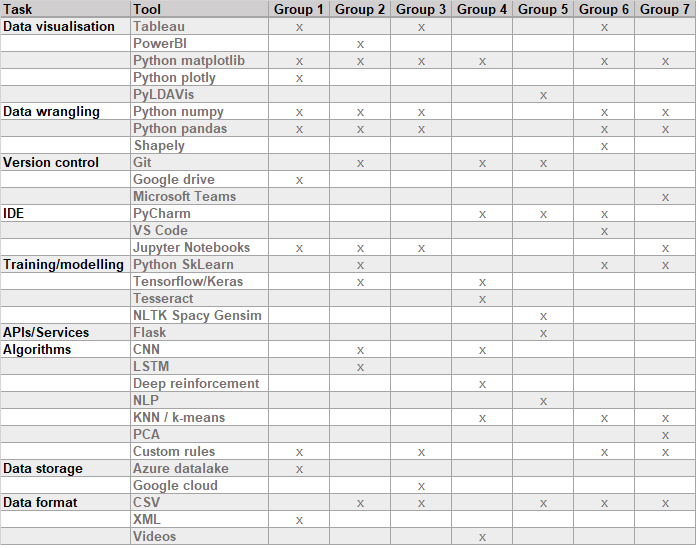}
\caption{Tools and techniques used in the group projects (not necessarily complete)}
\label{fig_tools}
\end{figure}

The group project interview was about the technical feasibility of the group project: did the students have the technical know-how to achieve valuable results for the customer with machine learning? The students confirmed this and what we saw from the final delivery to the customers also confirmed this. The only comment we got is that they would have liked to know more about testing of machine learning applications. An interesting side effect of the interviews is that it gave us an overview of the tools and techniques that were used in the group projects, see Figure \ref{fig_tools}. Note that Group 1 and Group 3 in the end did not use any machine learning algorithm to solve the customer problem. The list of tools used shows the technical background of our students: they use much more than is offered to them by the teachers. 

We sent an online survey to our graduation students, asking them to look back at their education. The main question was if they felt well-prepared for their graduation assignments. Five out of nine students had an NLP assignment, two had a reinforcement learning assignment, one a time series analysis assignment and one an image analysis assignment. This matches perfectly with the topics we included in Semester 7. All students are using Python for their graduation assignment. 

\begin{figure}
\includegraphics[width=\columnwidth]{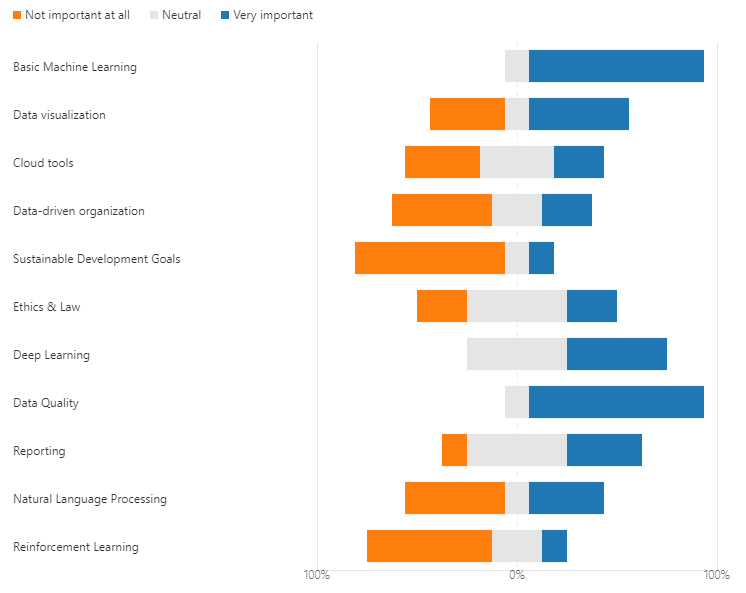}
\caption{Students estimation of the relevance of topics for their graduation assignment}
\label{fig_students}
\end{figure}

The graduation students collectively rated the topics of machine learning, deep learning and data quality as most relevant for their graduation assignment, see Figure \ref{fig_students}. The relevance of the other topics is more divided, due to the different nature of the assignments. Note that there are no topics that we included in our educational programme that score only in the categories ``Neutral'' or ``Not Important at All''.

Some quotes we got from our graduation students, that confirmed our idea that we are on the right track with our AI education:
\begin{quote}
    ``I like to work with large amounts of data with the aim of collecting new insights.''
\end{quote}
\begin{quote}
    ``I found it more interesting than regular software and I felt that my job opportunities would be better/more interesting''
\end{quote}
\begin{quote}
    ``I feel like you learn the most from the project. It allows you to discover and work with new technology, While also having the benefit that you are using the new technology in a realistic setting and you can cooperate with your fellow group members to achieve a better result.''
\end{quote}
\begin{quote}
``Got to really expand on my Data Science knowledge significantly because the way the course was structured - I got to try out stuff that I never knew/did in the past. Open programme allowed me to explore what I was personally interested into and the Development Report has been so far my favourite report to write (and I hate writing reports :D).''
\end{quote} 
\begin{quote}``I feel like my software development skills are essential within my assignment. For example multi threading to speed up cleaning etc. Something I learned during the Software courses.''
\end{quote} 
\begin{quote}
``In the market, different type of jobs exist in the area of Data Science. It can be somewhat vague. Most companies refer to them as type A (analyst) or B (builder). A lot of companies hire type A, but type B is more rare and becoming very important. The best fit for my profile (Software \& AI) is an AI engineer.''
\end{quote} 
\begin{quote}``Thanks to ICT \& AI I can fill in a specific niche that existed within my graduation company. A software developer with an affinity for data science. Allowing me to work both in the BI team and the software development team. ''
\end{quote} 
\begin{quote}
``I love it! Makes me feel like a wizard :D!''
\end{quote} 
Next to the above cited positive remarks we received a few negative remarks about organisational issues (schedule, communication, class rooms, etc.) or individual teachers, not about the structure of the programme.

The graduation students mentioned the following missing topics: 1) unsupervised learning and 2) deployment of a data science solution. Due to lack of study material, development time and resources these topics were not included yet in the first version of Semester 7. The students had to figure it out by themselves (with help of teacher experts of course) if it was needed for their project(s). These topics were first on our list to include in the base material for the second execution of the programme.   

Graduation supervisors, both from the graduation company and from our own university, confirmed that students were apt for the assignment. As one of the university supervisors states: ``sometimes the complexity of an assignment is not in creating ways to get value from data, but more in building a piece of data processing software.'' This confirms our findings: there is a need for professionals that are skilled in both software engineering, data engineering and machine learning. 

Company supervisors value the work of our students. One of them states: ``While working on the project, we've already gained a lot of insights in our own data.'' This is something we hear more often from organisations we work with in Semester 4, 7 and the internships. The only improvement that is suggested by the company supervisors is to include more data pre-processing best practices especially for natural language processing.

\subsection{Planned Improvements}
Next to the interviews, students also were invited for a group evaluation session with the teachers. Based on all feedback received we decided to implement the following improvements for September 2020 (the second edition of Semester 7):
\begin{itemize}
    \item We made minor updates to the workshops to include more hands-on tutorials, e.g. for Natural Language Processing and data preparation (stemming, vectorization, stop word removal, etc.). Our students typically learn by doing. 
    \item We have included the CRISP-DM manual \cite{CRISP} in the student material such that they have more elaborate information on how to execute each of the steps in the IBM data science methodology \cite{IBM}.
    \item We have added intermediate deliverables (report with code, e.g. Jupyter Notebook) for the group project such that teachers can better follow the group process. Moreover, this way the dataset and models can be handed over more easily to future project groups.
    \item We have hinted the students towards tooling that supports the data science process such as DVC and MLFlow, because they found out that Git alone is not enough for proper version control.
    \item We will provide the students with better guidance how to set up cloud data storage and cloud training. We have academic licenses for Azure and Google Cloud.
    \item We have added guidance for the students on how to use traditional ICT methods in a machine learning context, see \cite{Heck} and \cite{Heck3}.
    \item We have added reference material about ``missing topics'' such as unsupervised learning, deployment and testing of machine learning applications to the base material of Semester 7.  We will seek to invite guest lecturers on these topics during the semester.
\end{itemize}
We will not change anything in the educational concept of Semester 4 and 7, because we are satisfied with the results we had last time and we did not get any feedback that made us think otherwise.\\

\section{Discussion and Lessons Learned}
In this section we will discuss the above programme, our experience with it and the feedback we received, resulting in lessons learned that could benefit others educating AI engineers. 

After the first-year experiences with our new AI educational programme we realised that the IBM methodology\cite{IBM} is crucial for showing the students that the machine learning engineering process differs from the standard software engineering process \cite{Heck}. So from that moment on we made the methodology even more central to our semesters. Students also have to use it to organise their internships for example. When we put the IBM data science methodology more on the forefront of the semester project we also added several clear milestones to give both students and teachers a way to follow the process in a structured way: 1) Business Proposal, 2) Exploratory Data Report, 3) Data Quality Report, 4) Experiment Reports, 5) Final Report. See also the CRISP-DM guide\cite{CRISP} for a more detailed description of steps and deliverables.

\begin{center}\fbox{\parbox{8cm}{\textbf{Lesson Learned 1} The way of working for machine learning applications is different from traditional rule-based software\cite{Heck}. We need a separate project phasing for this. We use the IBM Data Science Methodology\cite{IBM} (based on CRISP-DM\cite{CRISP}) for that, see Figure \ref{fig_IBM}.  }}\end{center}

In discussions with both students, industry partners and teachers we realised that teaching the students already in the beginning of their studies (Semester 4) that machine learning starts with data and how much time it takes to make the data ready for machine learning, is the most important basis for successful machine learning projects later on. The students need to understand that if a machine learning model performs badly, it might be because they are using the wrong data examples to train the model. The data processing steps are also very well presented in the IBM Data Science Methodology (again, see Figure \ref{fig_IBM}) that we use. This IBM methodology is based on CRISP-DM\cite{CRISP}, which comes from the Data Mining community.  

\begin{center}\fbox{\parbox{8cm}{\textbf{Lesson Learned 2} Learning about data collection, data storage and data cleaning is a must; a considerable amount of time is spent on data preparation in many machine learning projects and this is different from software engineering projects.}}\end{center}

Related to this we also saw that in the beginning of machine learning projects time is spent to discuss the meaning and structure of the data with the customer, whereas in software engineering projects you would discuss software requirements. This requires the students to master data engineering and data understanding. For communicating data understanding, different types of data visualisations need to be made. This is a skill that is not included in our software engineering programme, thus we had to include it in our AI programme. We teach basic plots in a programming environment (Python's matplotlib library) but also the principles of story telling with data in interactive tools (Tableau).  

\begin{center}\fbox{\parbox{8cm}{\textbf{Lesson Learned 3} It is essential to dive into the domain of the client in order to understand the data and to improve the models. This implies close collaboration with domain experts and a way of communicating with them about the data and your models. Data visualisation is key for that.}}\end{center}

During our search for the state of the practice we found more and more tools that take-over the manual model training process from the engineer: you just upload some labelled data, and a smart black-box tool creates the most optimal model for you. And of course also the cognitive APIs where a pre-trained model is running and you just send your unlabelled data to receive the corresponding labels. Although we think the future of AI engineering would more and more make use of such Auto-ML solutions, we do feel it is crucial for the students to experience the process of manually training models, because only then they can understand the inner workings of such models on a conceptual level. Understanding the concepts of model training in our opinion is needed for preparing the right data and evaluating the fitness of the model outcome, even if you are not training the model yourself. An AI engineer needs to be able to assess if an auto-ML solution would fit the customer needs or if a custom ML solution is required.   

\begin{center}\fbox{\parbox{8cm}{\textbf{Lesson Learned 4} Even when automated machine learning tools like Auto-ML or cognitive API's are used, the experience of manually training the models on a given dataset should not be disregarded. If you only experience the easy 'inference' part of ML (which is like calling a function) you completely miss the point of creating, training, optimising models and evaluating their predictive quality. This makes you less apt in tailoring a machine learning solution to specific customer needs. }}\end{center}

One of our main concerns from the start of the programme was the fact that our population of students is not very proficient and not very interested in mathematics. Would it be possible to teach them machine learning without going too much into the mathematics behind it? As you can see we did not include any specific mathematics courses into our programme and up til now we do not have the feeling that we should. We explain all machine learning concepts on an intuitive level. We try to select the most basic variant of each concept and give pointers to more advanced versions for those students that want to learn more. Our expectation is that the majority of our AI engineers would work in an environment where master educated data scientists have the lead in designing models. As said before, we still think that the experience of training simple models yourself is very valuable in working together with those data scientists later on. We saw that even with these "simple models" our students manage to achieve valuable results in their own projects or for the external clients. 

\begin{center}\fbox{\parbox{8cm}{\textbf{Lesson Learned 5} An AI engineer does not need to learn advanced mathematics (linear algebra, calculus, probability theory, etc.) to become proficient in applying machine learning algorithms to solve practical problems.}}\end{center}

Another reason to limit the programme to the treatment of basic variants of concepts is the limited amount of time we have available (two semesters). The feedback we received, gives us the impression that the selection we made provides the students with a broad enough basis in machine learning to continue with deep-dives on their own during their internships or later career. We also are very happy with the open programme part of our semesters were students are challenged to chose their own deep-dive topics. It immediately shows them that there is much more to machine learning and it motivates them to reflect on their personal preferences in the field. 

\begin{center}\fbox{\parbox{8cm}{\textbf{Lesson Learned 6} For beginning AI engineers it is necessary to have some guidance on where to start with machine learning. The amount of online material is overwhelming. For our students we simplified it to a limited set of algorithms and validation methods, focusing on the basic concepts behind them. We offer students the opportunity to deep-dive on topics of their own choice.}}\end{center}

We are lucky to have a population of students that is educated as a software engineer and then combines this with two semesters in machine learning to become a complete AI engineer. The first author had the chance to start an applied research project into machine learning engineering to discover the difference with software engineering\cite{Heck} and collect material on e.g. machine learning testing\cite{Heck2}. We hugely benefited from this effort to determine the delta in skills between software engineers and AI engineers and shape the AI semesters based on that. As an AI engineer you need to accustom yourself to tooling that supports the data science process, next to tooling that supports the software engineering process. Tools that support both processes are relatively new or not existing yet. For example, for version control we recommend our students to use tools like DVC (dataset versioning) and MLFlow (model versioning) as an extension to Git (code versioning).     

\begin{center}\fbox{\parbox{8cm}{\textbf{Lesson Learned 7} A good AI engineer ought to combine knowledge on machine learning topics with knowledge on software engineering and devops: deployment of trained models, data pipelines, version control, testing, CI/CD, etc. You need to understand what is needed to deliver a production-ready software system that includes machine learning components.}}\end{center}

\subsection{Implications for practitioners}
All lessons learned highlighted above are also valid as best practice for practitioners working as AI engineer. We see software engineers in industry currently transitioning into a role of AI engineer. We plan to build some spin-off training programs for those practitioners based on what we have for our students. We also transfer our experience in AI engineering to the industry by supervising internship and graduation students who execute their projects in industry. And we plan to continue publishing our collected best practices for the industry community in the form of blog posts, conference talks, etc.

\section{Related work}
Turning software engineers into AI engineers starts by understanding the difference between the two. We found publications as far back as the 1980s on this topic, but only recently interest has been drawn again to this question with the widespread application of machine learning \cite{AI}. 

We have summarised what we found in two online posts. The first post is about the difference between traditional (rule-based) software engineering and engineering machine learning applications \cite{Heck}. The second post further elaborates on the specificity and importance of testing machine learning applications \cite{Heck2}. Both posts have ample references to related work. In the remainder of this paper we will include and highlight those references that are directly used for developing and implementing our ICT \& AI educational programme. 

Worth mentioning here is the book "Machine Learning Engineering" by Andriy Burkov (\url{http://www.mlebook.com/wiki/doku.php}). This book is still under development but discusses all steps in a machine learning engineering life cycle. Burkov shares our vision on machine learning engineering: "Machine learning engineering (MLE) is the use of scientific principles, tools, and techniques of machine learning and traditional software engineering to design and build complex computing systems. MLE encompasses all stages from data collection, to model building, to making the model available for use by the product or the customers." Our educational programme provides a practical implementation of all steps included in the book.

When designing our program we made use of the Edison Framework (\url{https://edison-project.eu/edison/edison-data-science-framework-edsf/}). The Edison Framework is a deliverable of an EU research project to establish a baseline for the upcoming profession of a data scientist. It defines three levels of the data science competences: associate, professional and expert level. In terms of the Edison Analytics competences our students are on associate level (due to a lack of mathematical background): basic or entry level that defines minimum competences and skills to be able to work in a data science team under supervision. For the Edison Engineering competences our students need to reach a professional level: this indicates the ability to solve major tasks independently, use multiple languages, tools and platforms and develop specialised applications. The Edison Framework also identifies a number of profiles. Our programme most closely matches that of Data Science Engineer: designs/develops/codes large data (science) analytics applications to support scientific or enterprise/business processes. We used the Edison Framework to check for missing topics and knowledge areas, but did not find any major gaps. The Edison Framework also helped us to frame our programme in relation to other data science programmes. 

In software engineering education conferences/tracks (ICSE-SEET, SIGCSE and ITiSCE) only recently the topic of AI engineering (also referred to as data science or machine learning) gained more attention. The 2020 SIGSCE conference included a separate track on "Data Science" for the first time. The 2020 ICSE-SEET track included a paper with detailed information on "Software Engineering for AI-enabled Systems (SE4AI)"\cite{SE4AI}. Note that this is a single course whereas we present an entire bachelor curriculum. The course (\url{https://ckaestne.github.io/seai/}) has similar objectives to our ICT \& AI programme, but is for master students. The advantage we have at Fontys UAS is that it is common practice for our students to work on practical assignments from real companies where SE4AI had to come up with a made-up classroom project to demonstrate the theory. This is rather hard in the field of machine learning where large datasets with hidden patterns are needed.

The problem of creating relevant assignments for data science is also discussed by Bhavya et al.\cite{Bhavya}. They follow our idea of creating projects that span entire semesters. However, they still use an artificial assignment (search engine development) constructed by the teachers, whereas we use semester long projects from external stakeholders. We complement this with teacher-guided assignments (the challenges) to make sure students have the opportunity to proof all required learning outcomes during the semester. 

The other reports we found in those software engineering education tracks were more focused on the mathematical side of AI (developing new algorithms) instead of on the engineering side (implementing AI in a software system). This confirms our idea that AI engineering is an emerging field. Currently, a considerable share of AI engineering projects are in the experimentation or prototyping stage and not yet working on what is needed to implement AI in a production-ready software system\cite{Bosch2}. The research agenda presented by Bosch et al.\cite{Bosch2} is in fact also an agenda for the future content of AI programmes.  

\section{Conclusion}
We started out with the question: how should we educate software engineers to become AI engineers. First of all, our experience made us realise that it is possible to do this. We now have bachelor level applied AI engineers that succeed in delivering AI solutions for a client, even with a limited mathematical background. They have the skills to execute the full cycle from business understanding till model deployment. 
This paper contributes in the following ways:
\begin{enumerate}
     \item We have confirmed that there is indeed a need for the AI engineer that combines software engineering skills with machine learning (and data!) skills. 
    \item We have described the structure, didactic concept and content of our educational programme on AI, that builds upon our educational programme on software engineering to educate AI engineers.
    \item We have illustrated the methods, tools, techniques and frameworks that we deem important for AI engineers. 
    \item We have summarised seven lessons learned that are also valuable for practitioners.
    \item We have verified with our students and their supervisors that they are indeed apt to fulfill their graduation assignment in an external company as AI engineer. 
\end{enumerate}
The educational programme that we have right now seems to work for our community of ICT students. We plan to verify if a similar programme could also work for practitioners switching from software engineer to AI engineer. And of course we will keep updating the content of our programme to the newest developments in the fast evolving field of AI engineering. 

\section*{Acknowledgement}
We would like to thank all teachers, students and company representatives that helped us in shaping our education and were kind enough to provide us with feedback on our programme. Special thanks go to Lu\'is Cruz from the Delft University of Technology for his valuable feedback on our paper and his help in exploring the field of AI engineering.

\vspace{12pt}
\end{document}